\documentclass[twocolumn,letterpaper,amsmath,amssymb,floatfix,prl, showpacs,superscriptaddress]{revtex4} 

\def\ul#1#2{\textstyle{\frac{#1}{#2}}}
\def\tr{{\mathrm{Tr}}}
\newcommand {\vct}[1] {\mathbf {#1}}

\usepackage{graphicx}
\usepackage{dcolumn}  
\usepackage{bm}  
\usepackage{epsfig}

\begin{document}
\setlength\arraycolsep{2pt} 

\title{Fluctuation-Induced Interaction between Randomly Charged Dielectrics}

\author{Ali Naji}
\affiliation{Department of Physics and Astronomy, University of Sheffield, Sheffield S3 7RH, United Kingdom}   
\affiliation{Kavli Institute of Theoretical Physics, University of California, Santa Barbara, CA 93106, USA}
\affiliation{Department of Physics, Department of Chemistry and Biochemistry, \& Materials Research Laboratory, 
University of California, Santa Barbara, CA 93106, USA}
\affiliation{School of Physics, Institute for Research in Fundamental Sciences (IPM), P.O. Box 19395-5531, Tehran, Iran}

\author{David S. Dean}
\affiliation{Kavli Institute of Theoretical Physics, University of California, Santa Barbara, CA 93106, USA}
\affiliation{Universit\'e de Toulouse, UPS, Laboratoire de Physique Th\'eorique (IRSAMC),  F-31062 Toulouse, France}

\author{Jalal Sarabadani}    
\affiliation{School of Physics, Institute for Research in Fundamental Sciences (IPM), P.O. Box 19395-5531, Tehran, Iran}
\affiliation{Department of Physics, University of Isfahan, Isfahan 81746, Iran}

\author{Ron R. Horgan}
\affiliation{Kavli Institute of Theoretical Physics, University of California, Santa Barbara, CA 93106, USA}
\affiliation{DAMTP, CMS, University of Cambridge, Cambridge CB3 0WA, United Kingdom}

\author{Rudolf Podgornik}   
\affiliation{Kavli Institute of Theoretical Physics, University of California, Santa Barbara, CA 93106, USA}
\affiliation{Department of Theoretical Physics, J. Stefan Institute, SI-1000 Ljubljana, Slovenia}
\affiliation{Institute of Biophysics, School of Medicine and Department of Physics, Faculty of Mathematics and Physics, University of Ljubljana, SI-1000 Ljubljana, Slovenia}

\begin{abstract}
Monopolar charge disorder effects are studied in the context of  
fluctuation-induced interactions between neutral dielectric slabs. 
It is shown  
that quenched bulk charge disorder gives rise to an 
additive contribution to the net interaction force which decays as the  
inverse distance between the slabs and may thus 
completely mask the standard  Casimir--van der Waals 
force at  large separations. 
By contrast, annealed (bulk or surface) charge disorder leads to a net 
interaction force whose large-distance behavior 
coincides with the universal Casimir force between perfect conductors,
which scales as inverse cubic distance, and the 
dielectric properties enter only in subleading corrections.  
\end{abstract}


\maketitle

Recent ultrahigh sensitivity experiments on Casimir (zero temperature and ideally polarizable surfaces) and van der Waals (finite temperature and non-ideally polarizable surfaces) interactions between surfaces {\em in vacuo} \cite{bordag,kim} have highlighted the need for an accurate assessment of the possible electrostatic contribution to the total interaction when the  surfaces bear a disordered charge distribution \cite{speake}. The surface charge distribution can have various origins. In the so-called patch effect,  the variation of the local crystallographic axes of the exposed surface  of a  clean polycrystalline sample can lead to a variation of the local surface potential \cite{barrett}. These variations are of course sample specific and depend heavily on the method of preparation of the samples. The electrostatic forces due to this surface potential disorder cannot be eliminated by grounding the two interacting surfaces. A similar type of surface charge disorder can also be expected  for amorphous films deposited on crystalline substrates. Surface annealing of these films  can produce a grain structure of an extent that can be larger than the thickness of the deposited surface film \cite{liu}. In addition, adsorption of various contaminants  can also influence the nature and type of the surface charge disorder.

\begin{figure}[b!]
\begin{center}
\vspace{-.5cm}
\includegraphics[angle=0,width=6cm]{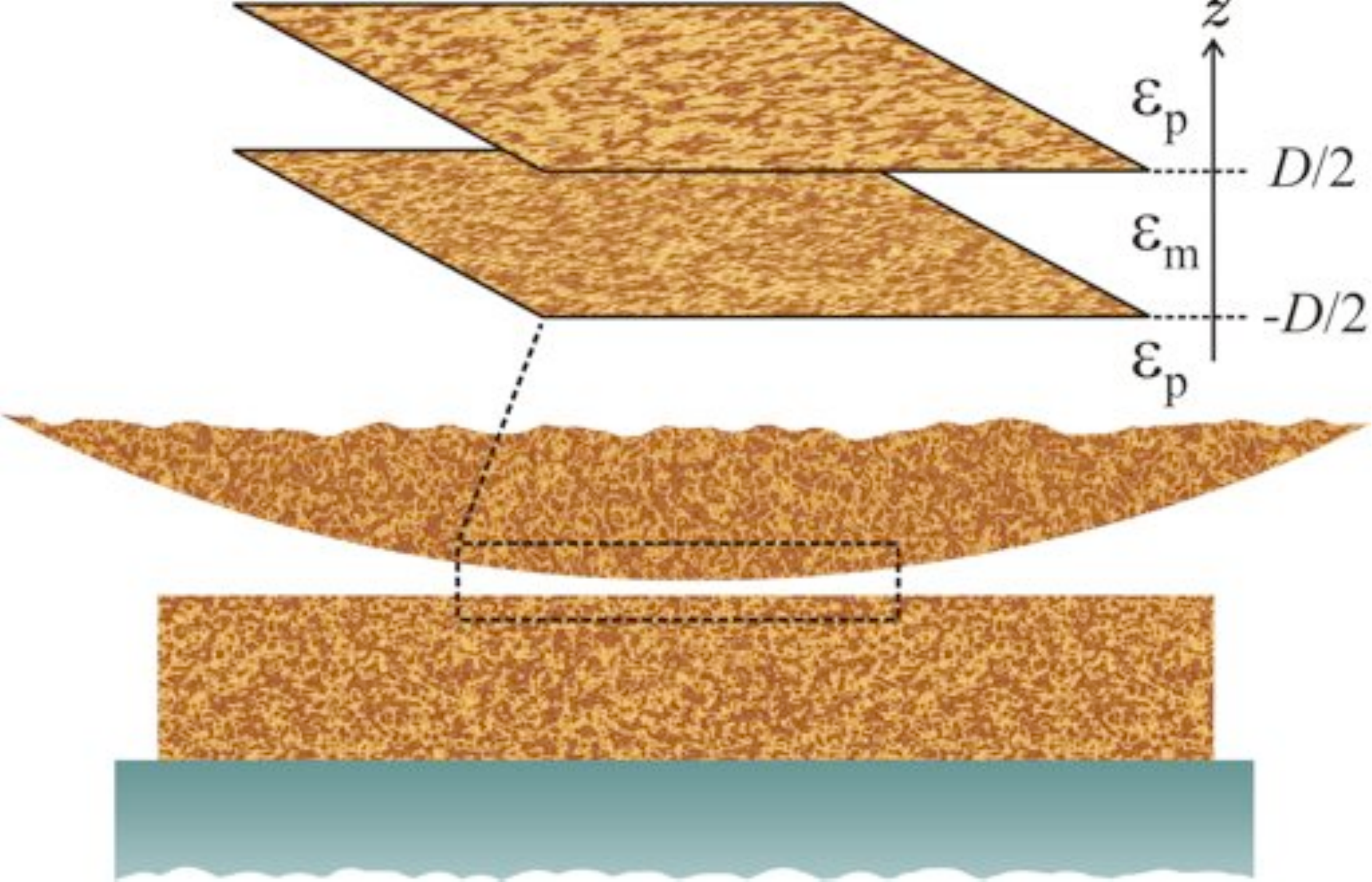}
\caption{(Color online) A typical experimental set up (bottom) is modeled with 
a plane-parallel system (top) of two dielectric slabs (half-spaces) 
of dielectric constant $\varepsilon_p$ interacting across a medium 
of dielectric constant $\varepsilon_m$. 
The charge distribution in the bulk of the slabs and on the two bounding 
surfaces at $z = \pm D/2$ has a disordered component (shown schematically by small light and dark patches)
with zero mean but finite variance, and may be either quenched or annealed in nature.}
\label{fig:schematic}
\vspace{-.5cm}
\end{center}
\end{figure}

Here we assess the effect of various types of {\em monopolar charge disorder}
 on the  interaction between two macroscopic  surfaces, delimiting two semi-infinite net-neutral dielectric slabs, separated by a layer of vacuum or an arbitrary dielectric material (Fig. \ref{fig:schematic}). 
Since the nature and distribution of the charge disorder in any of the experiments is in general not known, we consider different {\em a priori} models for the distribution  of disorder. Specifically, we assume that the charge disorder originates from  randomly distributed  monopolar charges which
may be present both in the bulk and on the bounding surfaces and can be either annealed or  
quenched.  It turns out that the type and the nature of the disorder has important consequences for the total interaction between apposed bodies and can even dominate or give a contribution comparable to  the underlying Casimir--van der Waals (vdW) effect. Our main goal is thus to investigate the {\em interaction fingerprint} of the charge disorder and to compare its contribution to the total interaction 
with the zero-frequency Casimir--vdW interaction between macroscopic surfaces.  This may in turn help in assessing whether the experimentally observed interactions can be interpreted in terms of disorder-induced effects  or pure Casimir--vdW interactions. 

We consider two semi-infinite slabs of dielectric constant $ \varepsilon_{p}$ and temperature $T$ with parallel planar inner surfaces (of infinite area $S$)
located normal to the $z$ axis
at $z=\pm D/2$,  where $D$ is thus the distance between their surfaces (see Fig. \ref{fig:schematic}).
The inner gap is filled with a material of dielectric constant $ \varepsilon_{m}$.
We shall assume that the dielectric slabs have a disordered monopolar 
charge distribution, $\rho(\vct r)$, which may arise from charges residing on bounding surfaces [$\rho_s(\vct r)$] and/or in the bulk [$\rho_b(\vct r)$], i.e. $\rho(\vct r)=\rho_s(\vct r)+\rho_b(\vct r)$. The charge disorder will be taken to be of zero mean (i.e., the slabs are net neutral) and  Gaussian-distributed  \cite{note0} with  no correlation in space {\em i.e.}, $\langle \!   \langle \rho({\bf r})\rho({\bf r}') \rangle   \! \rangle  =  g(\vct r)  \delta({\bf r}-{\bf r}')$ [where 
$\langle \!   \langle  \cdots\rangle  \! \rangle $ denotes the disorder average]. 
The total correlation is the  sum of the surface and bulk
 correlations  $g(\vct r) = g_s(\vct r)+g_b(\vct r)$. For the slab geometry considered here, the charge distribution is assumed to be statistically invariant in the plane of the dielectrics but with a variance dependent  on $z$ as
\begin{eqnarray}
 g_s(\vct r) &=& g_se_0^2 [\delta(z + D/2) +  \delta(z - D/2)], 
 \label{eq:g_s}
 \\ 
 g_b(\vct r) &=&  \left\{
 \begin{array}{ll}
 	 g_be_0^2 & \quad  |z|>D/2,\\
	 0      & \quad  |z|<D/2, 
 \end{array}
\right.
 \label{eq:g_b}
\end{eqnarray}
where $e_0$ is the elementary charge. It is worth mentioning how this sort of disorder distribution might arise. If the bulk material has charge impurities at the sites ${\vct r}_i$ distributed uniformly 
and independently with density $n_b$ and charges $q_i=\pm q_be_0$  with equal probability, then we clearly have $\rho_b({\bf r}) = \sum_i q_i \delta({\bf r}-{\bf r}_i) $ and find
$\langle \!   \langle \rho_b({\bf r}) \rangle   \! \rangle =0 $ and 
$\langle \!   \langle \rho_b({\bf r})\rho_b({\bf r}) \rangle   \! \rangle  = q_b^2 e_0^2n_b \delta({\bf r}-{\bf r}')$. We can thus make the correspondence $g_b= q_b^2n_b$. Similarly, one
can  make the correspondence $g_s= q_s^2n_s$ with $n_s$ being the surface density of impurity 
charges $\pm q_s$ on bounding surfaces. 

The partition function for the classical Casimir--vdW interaction (the zero-frequency Matsubara modes of the electromagnetic field)  may be written as
a functional integral over the  scalar field $\phi(\vct r)$,
\begin{equation}
{\mathcal Z}[\rho(\vct r)] = \int [{\mathcal D}\phi(\vct r)] \,\,e^{- \beta {\mathcal S}[\phi(\vct r); \rho(\vct r)]},
\label{eq:Z}
\end{equation}
with $ \beta = 1/k_{\mathrm{B}}T$ and  the effective action 
\begin{equation}
{\mathcal S}[\phi(\vct r); \rho(\vct r)] =  \int {\mathrm{d}} \vct r\, \big[{\ul{1}{2}}  \varepsilon_{0}
\varepsilon({\mathbf  r})\, (\nabla \phi(\vct r))^{2}   + {\mathrm{i}} \, \rho(\vct r) \phi(\vct r)\big], 
\label{act-1}
\end{equation}
where $\varepsilon({\mathbf  r}) = \varepsilon_p$ for $|z|>D/2$ and $\varepsilon_m$ otherwise. 
In order to evaluate the averaged quantities such as the effective interaction, one needs to average the partition function over different realizations of the disordered charge distribution, $ \rho(\vct r)$ \cite{ali-rudi}.  It is thus important to distinguish between {\em quenched} and {\em annealed}
 disorder that involve different averaging schemes. 
For quenched disorder, the disorder charges are frozen
and cannot fluctuate; for annealed disorder the charges 
can fluctuate and, in particular, the charge distribution in 
the two slabs can adapt itself to minimize the free energy 
of the system. These two different disorder types lead to very 
different physical behaviors as will be demonstrated below.

Let us first consider the {\em quenched disorder model}, where one must take the disorder average 
over the sample free energy, $\ln {\mathcal Z}[\rho(\vct r)]$, in order
to calculate the averaged quantities. The free energy of the quenched 
system, 
\begin{equation}
\label{eq:quenched_freen}
 \beta {\mathcal F}_{\mathrm{quenched}}
	= -  \langle \!   \langle  \ln {\mathcal Z}[\rho(\vct r)]  \rangle   \! \rangle, 
\end{equation}
may be calculated from Eq. (\ref{eq:Z}) as 
\begin{equation}
 \beta{\mathcal F}_{\mathrm{quenched}}  =
	      {\frac{1}{2}}\tr \ln G^{-1}({\vct r}, {\vct r}') + {\frac{ \beta}{2}} \tr \big\{ g({\vct r})G({\vct r}, {\vct r}') \big\},  
	   \label{eq:F_quenched}
\end{equation}
where $G({\vct r}, {\vct r}')$ is the Green's function defined via 
\begin{equation}
\varepsilon_0 \nabla \cdot [\varepsilon(\vct r) \nabla  G(\vct r, \vct r')] = -\delta(\vct r - \vct r'). 
\label{green1}
\end{equation}
In the first  term of Eq. (\ref{eq:F_quenched}), we recognize the standard 
zero-frequency vdW interaction $\beta {\mathcal F}_{\mathrm{vdW}}\equiv {\ul{1}{2}}\tr \ln G^{-1}({\vct r}, {\vct r}')$. The second term, $\beta {\mathcal F}_{g}\equiv {\ul{\beta}{2}} \tr \{g({\vct r})G({\vct r}, {\vct r}')\}$, represents the contribution of the quenched charge disorder, which 
turns out to be  additive in the free energy. 

\begin{figure*}[t!]
\begin{center}
\includegraphics[angle=0,width=15cm]{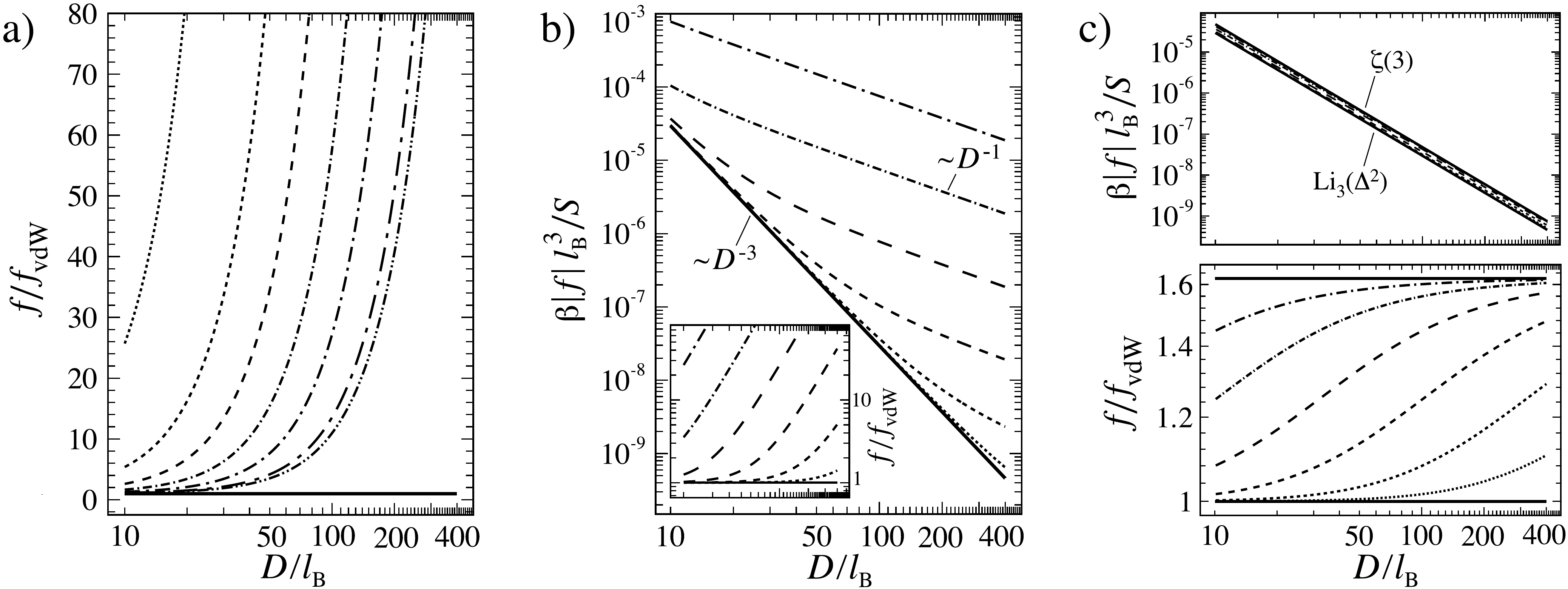}
\vspace{-.3cm}
\caption{a) Ratio of the total force (\ref{eq:total_f_quenched}) 
to the zero-frequency vdW 
force (\ref{eq:vdW_f})
between net-neutral dielectric half-spaces (in vacuum, 
$\varepsilon_m=1$) bearing quenched monopolar
charge disorder for fixed 
 bulk and surface disorder variances 
$g_b= 5\times 10^{-8}\,{\mathrm{nm}}^{-3}$,  
$g_s= \sqrt[3]{g_b^2}$
and different dielectric constants 
$\varepsilon_p=2, 5, 10, 20, 40, 80, 100$ (dashed curves from top).
b)  Magnitude of the rescaled total force 
(\ref{eq:total_f_quenched}) in the quenched case 
as a function of the rescaled distance 
for fixed $\varepsilon_p=10$, $g_s=0$ and various bulk disorder variances $g_b= 10^{-6}, 10^{-7}, 
10^{-8}, 10^{-9}, 10^{-10}, 10^{-11}\,{\mathrm{nm}}^{-3}$ 
(dashed curves from top). Solid curve is the pure vdW force 
 (\ref{eq:vdW_f}). Inset is the ratio of 
 the total force to the vdW force  (\ref{eq:vdW_f}) for the same range of $D$. 
 c) is the same as (b) but for   
annealed disorder (top panel, from Eq. (\ref{eq:F_fluct_charge_annealed})). Annealed curves stay 
close to one another and are bracketed by the perfect conductor result 
(Eq. (\ref{eq:metal}), top solid line, labeled by $\zeta(3)$) for large disorder 
and the vdW result  for no disorder (Eq. (\ref{eq:vdW_f}), bottom solid line, labeled by ${\mathrm{Li}}_3(\Delta^2)$) as seen more clearly from the
force ratio shown in the bottom panel.
(b) and (c) are plotted in log-log scale. 
}
\label{fig:forces}
\vspace{-.6cm}
\end{center}
\end{figure*}

The quenched expression  (\ref{eq:F_quenched}) is valid for any arbitrary disorder variance $g({\vct r})$. 
We now particularize to the case of planar dielectrics by using Eqs.  (\ref{eq:g_s}) and (\ref{eq:g_b}), in which case the zero-frequency vdW contribution 
per unit area,  
\begin{equation}
\frac{\beta {\mathcal F}_{\mathrm{vdW}}}{S} =  {\frac{1}{2}}\! \int\!\!\frac{{\mathrm{d}}^2Q}{(2\pi)^2} \ln{(1 - \Delta^{2}
\, e^{- 2 Q D})}, 
\label{eq:vdW}
\end{equation} 
yields the standard vdW force, $f_{\mathrm{vdW}} = -\partial {\mathcal F}_{\mathrm{vdW}}/\partial D$, as
\begin{equation}
\frac{\beta  f_{\mathrm{vdW}}}{S} = -\frac{{\mathrm{Li}}_3(\Delta^2)}{8\pi D^3}.
\label{eq:vdW_f}
\end{equation}
The  dielectric jump parameter  is defined as $\Delta = (\varepsilon_{p}  - \varepsilon_{m})/ (\varepsilon_{p}  + \varepsilon_{m})$ and $\mathrm{Li_3}(\cdot)$ is the trilogarithm function.
The bulk and surface disorder contributions 
are obtained as 
\begin{eqnarray}
\frac{ \beta {\mathcal F}_{g}}{S} &=&   -\frac{2g_bl_{\mathrm B} \varepsilon_m \Delta}{(\varepsilon_m+\varepsilon_p)^2}
\int_0^\infty \frac{{\mathrm{d}}Q}{Q} \frac{e^{-2QD}}{1 -\Delta^2e^{-2QD}} \nonumber\\
&& - \bigg(\frac{2g_sl_{\mathrm B}\varepsilon_{m} |\ln(1-\Delta^2)|}{
(\varepsilon_{m} +\varepsilon_{p})^2 \Delta}\bigg)\frac{1}{D}, 
\label{eq:F_g_quenched}
\end{eqnarray}  
at all separations $D$ with $l_{\mathrm B}=\beta e_0^2/(4\pi\varepsilon_0)\simeq 56.8$~nm being the Bjerrum length   in vacuum at room 
temperature ($T=300\,{\mathrm{K}}$). 
The quenched contribution from the bulk disorder (first term) in Eq. (\ref{eq:F_g_quenched})  is in  principle infra-red divergent, however the corresponding force is finite. The total force, $f_{\mathrm{quenched}} = -\partial {\mathcal F}_{\mathrm{quenched}}/\partial D$,  thus  follows as 
\begin{equation}
\frac{\beta  f_{\mathrm{quenched}}}{S} = -\frac{ g_b  l_{\mathrm B} \Delta}{2\varepsilon_pD} -  \bigg(\frac{2g_sl_{\mathrm B} \varepsilon_{m} |\ln(1-\Delta^2)|}{
(\varepsilon_{m} +\varepsilon_{p})^2 \Delta}\bigg)\frac{1}{D^2} -\frac{{\mathrm{Li}}_3(\Delta^2)}{8\pi D^3}. 
\label{eq:total_f_quenched}
\end{equation}
Here we obtain a sequence of scaling behaviors of different origins: a leading $1/D$ term due to the quenched bulk disorder, a subleading $1/D^2$ term  from the surface charge disorder, and the pure vdW term 
that goes as $1/D^3$ and prevails      
in the absence of charge disorder. 
The disorder contributions (first and second terms in Eq. (\ref{eq:total_f_quenched})) are attractive when the 
dielectric mismatch $\Delta>0$ (e.g., for 
two dielectrics slabs 
in vacuum) and repulsive otherwise 
(e.g., for the two surfaces of a single 
slab in vacuum). 
One might expect that globally electroneutral slabs would exhibit a dipolar-like interaction force
on the leading order rather than the monopolar forms $1/D$ (or $1/D^2$)  obtained for the bulk (or surface) charge
distribution. The physics involved is indeed subtle as the disorder terms result from the self-interaction
of the charges with their images (which follows from $G({\vct r}, {\vct r})$, Eq. (\ref{eq:F_quenched}),  
and only in a dielectrically inhomogeneous system) and not from dipolar interactions (which come from an expansion of $G({\vct r}, {\vct r}')$ 
when $|{\vct r}- {\vct r}'|$ is large). Statistically speaking each charge on average (as any other charge has an equal probability of being of
the same or opposite sign) only sees its image, thus explaining the leading monopolar form
in the net force.

The remarkable result is however the relative importance of the disorder-induced forces, which  exhibit 
a much weaker decay with the separation, $D$. They 
may thus completely mask the standard Casimir--vdW force at sufficiently large separations depending on the dielectric constants  
and the disorder variances as shown in Figs.~\ref{fig:forces}a and 
\ref{fig:forces}b; the bulk 
disorder variance is chosen
here within  the typical range 
$g_b\simeq 10^{-11}-10^{-6}\,{\mathrm{nm}}^{-3}$ 
(corresponding to impurity charge densities of 
$10^{10}-10^{15}\,e_0/{\mathrm{cm}}^{3}$) \cite{Kao_Pitaevskii}. 
 
For relatively small surface disorder ($g_s \ll g_b l_{\mathrm B}$), the anomalous $1/D$ behavior is predicted to dominate the
vdW $1/D^3$ behavior beyond the crossover distance
 \begin{equation}
 D_b= \left[\frac{\varepsilon_{p} {\mathrm{Li}}_3(\Delta^2)}{4\pi g_b l_{\mathrm B} \Delta}\right]^{1/2}, 
\end{equation}
which, for typical parameter values, covers
the experimentally relevant range of distances 
from a few hundreds of nm to several microns (Fig.~\ref{fig:forces}b, main set).
For strong surface disorder ($g_s \gg g_b l_{\mathrm B}$), on the 
other hand, one expects the
 $1/D^2$ behavior to dominate beyond 
\begin{equation}
 D_s  =  \frac{
(\varepsilon_{m} +\varepsilon_{p})^2  \Delta\, {\mathrm{Li}}_3(\Delta^2)}{16\pi g_sl_{\mathrm B} \varepsilon_{m}  |\ln(1-\Delta^2)|}. 
\end{equation} 

So far we have only examined the effects from the quenched disorder. In reality one may encounter disordered charges with some degree of annealing. A possible idealization is to assume that the disorder is completely annealed  (the intermediate partially annealed cases are also analytically tractable as shown recently \cite{mama} but will not be considered here).  In the {\em annealed disorder model}, the disorder average is taken over the sample partition function, $ {\mathcal Z}[\rho(\vct r)]$; hence, the free energy of the system, 
\begin{equation}
\label{eq:annealed_freen}
 \beta {\mathcal F}_{\mathrm{annealed}} 
	= -  \ln\, \langle \!   \langle  {\mathcal Z}[\rho(\vct r)]  \rangle   \! \rangle, 
\end{equation}
may be evaluated as 
\begin{equation}
\beta{\mathcal F}_{\mathrm{annealed}}  = {\frac{1}{2}}\tr \ln \big[G^{-1}({\vct r}, {\vct r}') + \beta g({\vct r}) \delta({\vct r}- {\vct r}') \big].
				\label{eq:F_annealed}			
\end{equation}
Note that, unlike  the quenched result in Eq. (\ref{eq:F_quenched}), 
the disorder and the pure Casimir--vdW 
contributions can not be separated in general when the disorder is annealed.  

In the case of two interacting slabs with the surface and bulk disorder variances (\ref{eq:g_s}) and (\ref{eq:g_b}), the modified inverse Green's function $G^{-1}({\vct r}, {\vct r}') + \beta g({\vct r}) \delta({\vct r}- {\vct r}')$ may be evaluated explicitly and the fluctuational trace-log term may be calculated by the standard methods \cite{rudi_jcp} as
\begin{equation}
	\frac{\beta {\mathcal F}_{\mathrm{annealed}} }{S}  = { \frac{1}{2} } \int \frac{{\mathrm{d}}^2Q}{(2\pi)^2} \ln \big(1-\Delta_g^2(Q)\,e^{-2QD}\big)
	\label{eq:F_fluct_charge_annealed}
\end{equation}
with 
\begin{equation}
\Delta_g(Q) = \frac{\varepsilon_{m} Q - \varepsilon_{p} \sqrt{Q^2 + 4\pi l_{\mathrm B} g_b/\varepsilon_{p}  } - 4\pi l_{\mathrm B} g_s}
{\varepsilon_{m} Q + \varepsilon_{p} \sqrt{Q^2 + 4\pi l_{\mathrm B} g_b/\varepsilon_{p}  } + 4\pi l_{\mathrm B} g_s}. 
\end{equation}
Let us first consider the large-distance behavior of the net annealed force. 
For strong annealed bulk disorder ($g_s \ll g_b l_{\mathrm B}$),  we obtain
 the asymptotic behavior 
\begin{equation}
	\frac{\beta f_{\mathrm{annealed}} }{S} \simeq 
-\cfrac{\zeta(3)}{8\pi D^3} + \bigg(\cfrac{ 3 \zeta(3) \varepsilon_m}{\sqrt{ 64\pi^3 g_b l_{\mathrm B} \varepsilon_p}}\bigg)\cfrac{1}{D^4}, 
		    \label{eq:total_f_annealed_1} 
\end{equation}	  
which is expected to hold for $D\gg 3\varepsilon_m/\sqrt{\pi g_b l_{\mathrm B} \varepsilon_p}$  [i.e., $D\gg 70$~nm for $g_b=10^{-6}\,{\mathrm{nm}}^{-3}$ and $\varepsilon_{p}= 10$ in vacuum]. 
While for weak bulk
 disorder ($g_s \gg g_b l_{\mathrm B}$),  we obtain
\begin{equation}
	\frac{\beta f_{\mathrm{annealed}} }{S} \simeq 
			-\cfrac{\zeta(3)}{8\pi D^3} + \bigg(\cfrac{3 \zeta(3) \varepsilon_m }{16\pi^2 g_sl_{\mathrm B}}\bigg)\cfrac{1}{D^4},  
			    \label{eq:total_f_annealed_2} 
\end{equation}	  
which is expected to hold for $D\gg 3\varepsilon_m /(2\pi g_s l_{\mathrm B})$  [i.e., $D\gg 80$~nm
for $g_s=10^{-4}\,{\mathrm{nm}}^{-2}$ in vacuum]. 
Obviously, material properties disappear in the leading-order {\em total} force between {\em arbitrary dielectrics} bearing annealed
charge disorder and one ends up with the universal attraction as
one would 
expect for two  perfect conductors 
\begin{equation}
	\frac{\beta f_{\mathrm{annealed}} }{S} = -\frac{\zeta(3)}{8\pi D^3} \qquad \quad D\rightarrow \infty. 
	    \label{eq:metal} 
\end{equation}	  
These asymptotic behaviors are also obtained for strong disorder ($g_b$ or $g_s\rightarrow \infty$). On the contrary, for weak disorder ($g_b$ and $g_s\rightarrow 0$) or for vanishing separation, one recovers the non-universal vdW force  (\ref{eq:vdW_f}) as the asymptotic behavior. 
It is thus interesting to note that the force 
in the annealed case is bounded between these two limiting results, i.e. Eqs. (\ref{eq:vdW_f}) and (\ref{eq:metal}) 
(Fig. \ref{fig:forces}c, solid lines). The above results demonstrate 
the intuitive fact that dielectric slabs with annealed charges tend to behave asymptotically as perfect conductors and, 
unlike the quenched case, the algebraic 
decay of the leading contribution to the net force remains unchanged.
The deviations due to material properties and 
the disorder variance contribute a repulsive 
subleading force (Eqs. (\ref{eq:total_f_annealed_1}) and (\ref{eq:total_f_annealed_2})). 

For the experimental {\em sphere-plane} geometry \cite{kim} a naive application of the {\em proximity force approximation} \cite{bordag} to the results derived above would lead to forces with the leading behavior $\sim \ln{D} + {\cal O}(D^{-1})$ in the quenched case (from Eq. (\ref{eq:total_f_quenched})) 
and $\sim D^{-2} + {\cal O}(D^{-3})$ in the annealed case (from 
Eqs. (\ref{eq:total_f_annealed_1})  and (\ref{eq:total_f_annealed_2})). 
Thus, an effective scaling exponent (defined as $D^{-\alpha}$) of  
$\alpha \leq 1$ (consistent with recent experimental observation of a residual electrostatic force scaling as $D^{-0.8}$  \cite{kim-new})  
may be obtained in the quenched case, both with the bulk disorder 
(plane-plane and sphere-plane geometry) and the surface disorder model
(sphere-plane geometry). A more detailed comparison with force measurements should be attempted once the experimental and methodological uncertainties surrounding experiments are sorted out (see Ref. \cite{kim} and published comments).

In conclusion, we have studied the influence of 
charge disorder on the fluctuation-induced 
interaction between net-neutral dielectric slabs bearing random 
quenched and/or annealed {\em monopolar charges} on their bounding surfaces
and/or in the bulk and compared it with the zero-frequency 
Casimir--vdW interaction between them. 
Quenched disorder leads to an additive contribution to the total 
interaction force that  scales as $1/D$ (or $1/D^2$) for bulk (or surface) charge disorder, may be attractive or repulsive and depends on the dielectric  constants of the materials. By contrast, annealed disorder leads to a net attractive force, which is universal  and decays
as $1/D^3$ at large separations. Thus, the main fingerprint of the annealed disorder is that 
the total force remains intact in this case as the dielectric constants
are varied. This could help distinguish this type of interaction from the pure Casimir-vdW interaction between
dielectrics with no disorder as well as from the interaction due to the quenched 
disorder, Eq. (\ref{eq:total_f_quenched}).  These two latter cases can in turn be distinguished  by monitoring the dependence on the separation of the net interaction, which for the quenched disorder exhibits a much weaker decay.
Note that the disorder effects are compared here with 
the zero-frequency Casimir-vdW interaction. The 
precise correction presented by the higher-order Matsubara 
frequencies is very material specific, but its magnitude
(relative to the zero-frequency term) is typically small for 
the most part of the separation range considered here 
and remains negligible in comparison with the quenched terms in (11). 

We thank M. Kardar and W.J. Kim for useful discussions. 
This research was supported in part by the 
NSF (Grant No. PHY05-51164).
D.S.D. acknowledges support from the Institut Universitaire de France. R.P. acknowledges support from ARRS.
A.N. is a Newton International Fellow.

\end{document}